\documentclass[aps,prd,onecolumn,showpacs,showkeys,amsmath,amssymb]{revtex4}
\usepackage{amsfonts}
\usepackage{amsmath}
\usepackage{graphicx}
\usepackage{subfigure}
\usepackage{dcolumn}
\usepackage{bm}
\usepackage{booktabs}
\usepackage{color}

\makeatletter
\newcommand{\figcaption}{\def\@captype{figure}\caption}
\newcommand{\tabcaption}{\def\@captype{table}\caption}

\newcommand{\Rmnum}[1]{\expandafter\@slowromancap\romannumeral #1@}
\def\hlinewd#1{%
  \noalign{\ifnum0=`}\fi\hrule \@height #1 \futurelet
   \reserved@a\@xhline}
\makeatother
\def\dab{\int^{\alpha_{max}}_{\alpha_{min}}d\alpha\int^{\beta_{max}}_{\beta_{min}}d\beta}
\def\uu{\langle\bar uu\rangle}
\def\dd{\langle\bar dd\rangle}
\def\qq{\langle\bar qq\rangle}
\def\GGa{\langle GG\rangle}
\def\GGb{\langle g_s^2GG\rangle}

\def\uGua{\langle\bar uGu\rangle}
\def\dGda{\langle\bar dGd\rangle}
\def\qGqa{\langle\bar qGq\rangle}
\def\qGqb{\langle\bar qg_s\sigma\cdot Gq\rangle}

\def\f(s){\left[(\alpha+\beta)m_Q^2-\alpha\beta s\right]}
\def\non{\\ \nonumber}

\usepackage{ulem}

\begin{document}

\title{Mass spectra of $Z_c$ and $Z_b$ exotic states as hadron molecules}

\author{Wei Chen}
\affiliation{Department of Physics
and Engineering Physics, University of Saskatchewan, Saskatoon, SK, S7N 5E2, Canada}
\author{T. G. Steele}
\affiliation{Department of Physics
and Engineering Physics, University of Saskatchewan, Saskatoon, SK, S7N 5E2, Canada}
\author{Hua-Xing Chen}
\email{hxchen@buaa.edu.cn}
\affiliation{School of Physics and Nuclear Energy Engineering and International Research Center for Nuclei and Particles in the Cosmos, Beihang University,
Beijing 100191, China}
\author{Shi-Lin Zhu}
\email{zhusl@pku.edu.cn}
\affiliation{School of Physics and State Key Laboratory of Nuclear Physics and Technology, Peking University, Beijing 100871, China \\
Collaborative Innovation Center of Quantum Matter, Beijing 100871, China \\
Center of High Energy Physics, Peking University, Beijing 100871, China
}

\begin{abstract}
We construct charmonium-like and bottomonium-like molecular interpolating currents with quantum numbers $J^{PC}=1^{+-}$ in a systematic way, including both
color singlet-singlet and color octet-octet structures. Using these interpolating currents, we calculate two-point correlation functions and perform QCD sum
rule analyses to obtain mass spectra of the charmonium-like and bottomonium-like molecular states. Masses of the charmonium-like $\bar qc\bar cq$ molecular
states for these various currents are extracted in the range 3.85--4.22~GeV, which are in good agreement with observed masses of the $Z_c$ resonances. Our numerical results suggest a possible landscape
of hadronic molecule interpretations of the newly-observed $Z_c$ states. Mass spectra of the bottomonium-like $\bar qb\bar bq$ molecular states are similarly obtained in the range 9.92-10.48~GeV, which support the interpretation of the $Z_b(10610)$ meson as a molecular state within theoretical uncertainties. Possible decay channels of these
molecular states are also discussed.
\end{abstract}



\maketitle



 \section{Introduction}\label{sec:intro}

To date, there are eight members in the family of the electrically charged states: $Z_c(3900)$, $Z_c(4020)$, $Z_1(4050)$, $Z_2(4250)$, $Z_c(4200)$, $Z(4430)$ and $Z_b(10610)$,
$Z_b(10650)$ observed in decays into final states containing a pair of heavy quarks~\cite{2008-Choi-p142001-142001,2014-Aaij-p222002-222002,2008-Mizuk-p72004-72004,2013-Ablikim-p252001-252001,2013-Liu-p252002-252002,2013-Xiao-p366-370,2013-Ablikim-p242001-242001,2014-Ablikim-p132001-132001,2014-Chilikin-p112009-112009,2014-Wang-p-,2011-Adachi-p-}.
Being not conventional $Q\bar Q$ states because of their charge, they must be exotic with minimal quark
contents $Q\bar Qu\bar d$.

The first charged exotic state, $Z(4430)^+$ was observed in the $B$ meson decay process $\bar B^0\to \psi(2S)\pi^+K^-$ by the Belle Collaboration \cite{2008-Choi-p142001-142001}
in 2007. Recently, the LHCb experiment repeated the Belle analysis and confirmed the existence of $Z(4430)^+$ with $J^P=1^+$ \cite{2014-Aaij-p222002-222002}. The broad doubly peaked
structure $Z_1(4050)^+$ and $Z_2(4250)^+$ are resonances in the $\pi^+\chi_{c1}$ channel, which were found by the Belle Collaboration \cite{2008-Mizuk-p72004-72004} in 2008. In 2013, the
BESIII Collaboration reported $Z_c(3900)^+$ in the process of $Y(4260)\to J/\psi\pi^+\pi^-$ \cite{2013-Ablikim-p252001-252001}, which was confirmed later by Belle \cite{2013-Liu-p252002-252002} and CLEO data \cite{2013-Xiao-p366-370}. The BESIII Collaboration also observed the $Z_c(4020)$ resonance in the $e^+e^-\to \pi^+\pi^-h_c$ and
$e^+e^-\to (D^{\ast}\bar D^{\ast})^{\pm}\pi^{\mp}$ processes \cite{2013-Ablikim-p242001-242001,2014-Ablikim-p132001-132001}. Very recently, two new charged charmonium-like resonances
$Z_c(4200)^+$ \cite{2014-Chilikin-p112009-112009} and $Z_c(4050)$ \cite{2014-Wang-p-} were observed by the Belle Collaboration in the processes of $\bar B^0\to J\psi\pi^-K^+$ and
$e^+e^-\to \pi^+\pi^-\psi(2S)$, respectively. For the charged bottomonium-like states, $Z_b(10610)$ and $Z_b(10650)$ were reported by the Belle Collaboration in the $\pi^{\pm}\Upsilon(nS)$
and $h_b\pi^{\pm}$ mass spectra in the $\Upsilon(5S)$ decay \cite{2011-Adachi-p-}. One can consult Refs.~\cite{2015-Esposito-p1530002-1530002,2015-Olsen-p101401-101401,Liu:2013waa} for recent reviews
of these charged resonances.

These exotic charged resonances are isovector states with 
quantum numbers $I^GJ^P=1^+1^{+}$ while their neutral partners have charge-conjugation parity $C=-1$.
As four-quark states with quark contents $c\bar cu\bar d$/$b\bar bu\bar d$, these newly observed resonances were usually studied as hadron molecules and tetraquark states. These two
hadron configurations are totally different. At the hadronic level, the hadron molecules are loosely bound states of two heavy mesons formed by the exchange of long-range color-singlet
mesons. Tetraquarks are more compact four-body states which are generally bound by the QCD colored force between diquarks at the quark-gluon level. There are many theoretical studies on
these charged resonances; see Ref.~\cite{Liu:2013waa} for a recent review. $Z_c(3900)^+$ was interpreted as a $\bar DD^*$ molecular state in
Refs.~\cite{2013-Wang-p132003-132003,2014-Aceti-p16003-16003,2014-Zhao-p94026-94026}. $Z_c(4020)^+$ was speculated
to be a $D^*\bar D^*$ molecular state in Refs. \cite{2014-Chen-p2773-2773,2014-Wang-p2761-2761,2013-Cui-p-}. $Z_c(4200)^+$ is much broader than other charged resonances in this family so that it was studied as a good candidate for a tetraquark state in Refs.~\cite{2014-Zhao-p94031-94031,2015-Chen-p-}. It was also studied as a molecular state in
Ref. \cite{2015-Wang-p-}. $Z(4430)^+$ was described as a tetraquark state in Refs.~\cite{2009-Bracco-p240-244,2008-Maiani-p73004-73004,2014-Maiani-p114010-114010} and a $D^{\ast}\bar D_1$ molecular state in Refs.~\cite{2008-Lee-p28-32,2007-Meng-p-,2007-Ding-p-}. $Z_b(10610)$ and $Z_b(10650)$ were interpreted
as $\bar BB^{\ast}$ and $\bar B^{\ast}B^{\ast}$ molecular states in Refs.~\cite{2011-Zhang-p312-315,2011-Sun-p54002-54002}.

The mass spectra of the charmonium-like and bottomonium-like tetraquark states were studied comprehensively in Refs. \cite{2006-Ebert-p214-219,2007-Ebert-p114015-114015,2010-Chen-p105018-105018,2011-Chen-p34010-34010,2013-Du-p33104-33104}. In Ref. \cite{2011-Chen-p34010-34010}, the masses of the charmonium-like tetraquark states with $I^GJ^{PC}=1^+1^{+-}$ were obtained from various currents in the range
4.0--4.2~GeV, which were consistent with the spectra of the charged $Z_c$ states. However, the molecular interpretations for these states are slightly more natural, especially for the $Z_c(3900)^+$ and $Z_c(4020)^+$ mesons which lie very close to the open-charm thresholds. In this work, we will study the mass spectra of the charmonium-like($\bar qc\bar cq$) and bottomonium-like($\bar qb\bar bq$) molecular states with the quantum numbers $J^{PC}=1^{+-}$ using the approach of QCD sum rules \cite{1979-Shifman-p385-447,1985-Reinders-p1-1,2000-Colangelo-p1495-1576}, which is used to study the hadron properties of the lowest bound state. The masses of higher
excited states are not easy to be calculate in QCD sum rules because their contributions are exponentially suppressed. However, there have been some attempts to study the orbitally excited nucleon~\cite{Jido:1996ia,Kondo:2005ur,Ohtani:2012ps}. We try to explain the newly observed $Z_c$ and $Z_b$ states as molecular states and compare the difference
between the molecular and tetraquark configurations. We note that we shall consider both the color singlet-singlet molecular structure and the color octet-octet ``molecular'' structure.

There are many investigations of possible molecular states by using hadronic level Feynman diagrams, particularly investigations in the framework of
the one boson exchange model \cite{2014-Zhao-p94026-94026}. Generally speaking, this kind of study employs an effective Lagrangian to derive either the scattering amplitude or the
effective potential. Besides the pion, quite a few other mesons are introduced which lead to many new coupling constants which have not been completely determined experimentally.
Moreover, a form factor is always introduced at each vertex in order to suppress the high momentum exchange effect, which requires a new cutoff parameter. In other words, there exits
some inherent uncertainties with the approach at the hadronic level. The QCD sum rule approach and the formalism at the hadronic level are complementary to each other.

The paper is organized as follows. In Sect.~\Rmnum{2}, we construct the molecular interpolating currents with $J^{PC}=1^{+-}$ for the $Z_c$ and $Z_b$ states. In Sect.~\Rmnum{3}, we
introduce the QCD sum rule formalism concisely and calculate the two-point correlation functions and spectral densities using these interpolating currents. We perform numerical analyses
and extract the mass spectra and coupling constants for the charmonium-like and bottomonium-like molecular states in Sect.~\Rmnum{4}. In the last section, we summarize our results and
discuss the possible decay modes for these molecular states.

\section{Molecular interpolating currents for the $Z_c$ and $Z_b$ states}\label{Sec:Currents}

There are two different types of four-quark operators: diquark-antidiquark type tetraquark fields and meson-meson type molecule fields.
The former kind of operator is composed of a pair of diquark and antidiquark fields ($[qq][\bar q\bar q]$) while the latter one is composed
of a pair of meson (or meson-like) fields ($[q\bar q][q\bar q]$). The diquark-antidiquark type tetraquark fields have been constructed and studied systematically in Refs.~\cite{2010-Chen-p105018-105018,2011-Chen-p34010-34010,2013-Du-p33104-33104}. Particularly,
there are eight independent diquark-antidiquark tetraquark fields with $J^{PC}=1^{+-}$, which have been systematically constructed and studied
in Ref.~\cite{2011-Chen-p34010-34010}. These eight
tetraquark currents can be transformed into the combinations of other eight meson-meson type molecule currents by using Fierz transformations.
In this paper, we systematically construct these eight meson-meson type molecule fields, and use them to study the charged $Z_c^+$ resonances as molecular states.

The color structure of a molecule field($[q\bar Q][Q\bar q]$) can be written as
\begin{eqnarray}
\nonumber
\left(\mathbf{3} \otimes \mathbf{\bar 3}\right)_{[q\bar Q]} \otimes \left(\mathbf{3} \otimes \mathbf{\bar{3}}\right)_{[Q\bar q]}
&=& \left( \mathbf{1}\oplus\mathbf{8} \right)_{[q\bar Q]} \otimes \left( \mathbf{1}\oplus\mathbf{8} \right)_{[Q\bar q]}
\non &=& \left( \mathbf{1}\otimes\mathbf{1} \right)\oplus \left( \mathbf{1}\otimes\mathbf{8}\right) \oplus \left( \mathbf{8}\otimes\mathbf{1}\right) \oplus
\left( \mathbf{8}\otimes\mathbf{8} \right)
\\ &=& \mathbf{1}\oplus \mathbf{8}\oplus\mathbf{8}\oplus\left(\mathbf{1}\oplus \mathbf{8}\oplus \mathbf{8}\oplus \mathbf{10}\oplus
\mathbf{\overline{10}}\oplus\mathbf{27}\right) \, . \label{eq:color}
\end{eqnarray}
The two color singlet structures in Eq.~\eqref{eq:color} come from the $\left(\mathbf{1}_{[q\bar Q]}\otimes\mathbf{1}_{[Q\bar q]}\right)$
and $\left(\mathbf{8}_{[q\bar Q]}\otimes\mathbf{8}_{[Q\bar q]}\right)$ terms in the second step, respectively. In other words, the two mesonic fields $[q\bar Q]$
and $[Q\bar q]$ should have the same color structures to compose a color singlet molecular current.

For $\left(\mathbf{1}_{[q\bar Q]}\otimes\mathbf{1}_{[Q\bar q]}\right)$ structure, we can obtain
eight independent molecule fields with $J^P=1^+$ by considering only S-wave of the angular momentum between the two mesonic fields.
Four of them are
\begin{eqnarray}
\nonumber J_{1\mu,L}^{(\mathbf 1)}&=&(\bar q_a\gamma_5Q_a)(\bar{Q}_b\gamma_{\mu}q_b),
\\ \nonumber J_{2\mu,L}^{(\mathbf 1)}&=&(\bar q_aQ_a)(\bar{Q}_b\gamma_{\mu}\gamma_5q_b),
\\
J_{3\mu,L}^{(\mathbf 1)}&=&(\bar q_a\gamma^{\alpha}Q_a)(\bar{Q}_b\sigma_{\alpha\mu}\gamma_5q_b), \label{eq:11L}
\\ \nonumber
J_{4\mu,L}^{(\mathbf 1)}&=&(\bar q_a\gamma^{\alpha}\gamma_5Q_a)(\bar{Q}_b\sigma_{\alpha\mu}q_b),
\end{eqnarray}
while the other four can be obtained by performing the charge conjugation transform
to these operators:
\begin{eqnarray}
\nonumber J_{1\mu,R}^{(\mathbf 1)}&=&-(\bar q_a\gamma_{\mu}Q_a)(\bar{Q}_b\gamma_5q_b),
\\ \nonumber J_{2\mu,R}^{(\mathbf 1)}&=&(\bar q_a\gamma_{\mu}\gamma_5Q_a)(\bar{Q}_bq_b),
\\
J_{3\mu,R}^{(\mathbf 1)}&=&(\bar q_a\sigma_{\alpha\mu}\gamma_5Q_a)(\bar{Q}_b\gamma^{\alpha}q_b), \label{eq:11R}
\\ \nonumber
J_{4\mu,R}^{(\mathbf 1)}&=&-(\bar q_a\sigma_{\alpha\mu}Q_a)(\bar{Q}_b\gamma^{\alpha}\gamma_5q_b).
\end{eqnarray}
In these expressions the subscripts $a$ and $b$ are color indices, and $q$ and $Q$ represent light quarks($u, d, s$) and heavy quarks($c, b$), respectively.

Similarly, we can construct eight independent molecule fields belonging to $\left(\mathbf{8}_{[q\bar Q]}\otimes\mathbf{8}_{[Q\bar q]}\right)$ color structure. Four of them are
\begin{eqnarray}
\nonumber J_{1\mu,L}^{(\mathbf 8)}&=&(\bar q_a\gamma_5\lambda^n_{ab} Q_b)(\bar{Q}_c\gamma_{\mu}\lambda^n_{cd}q_d),
\\ \nonumber J_{2\mu,L}^{(\mathbf 8)}&=&(\bar q_a\lambda^n_{ab}Q_b)(\bar{Q}_c\gamma_{\mu}\gamma_5\lambda^n_{cd}q_d),
\\
J_{3\mu,L}^{(\mathbf 8)}&=&(\bar q_a\gamma^{\alpha}\lambda^n_{ab}Q_b)(\bar{Q}_c\sigma_{\alpha\mu}\gamma_5\lambda^n_{cd}q_d), \label{eq:88L}
\\ \nonumber
J_{4\mu,L}^{(\mathbf 8)}&=&(\bar q_a\gamma^{\alpha}\gamma_5\lambda^n_{ab}Q_b)(\bar{Q}_c\sigma_{\alpha\mu}\lambda^n_{cd}q_d),
\end{eqnarray}
while the other four can be similarly obtained by performing the charge conjugation transform
to these operators:
\begin{eqnarray}
\nonumber J_{1\mu,R}^{(\mathbf 8)}&=&-(\bar q_a\gamma_{\mu}\lambda^n_{ab}Q_b)(\bar{Q}_c\gamma_5\lambda^n_{cd}q_d),
\\ \nonumber J_{2\mu,R}^{(\mathbf 8)}&=&(\bar q_a\gamma_{\mu}\gamma_5\lambda^n_{ab}Q_b)(\bar{Q}_c\lambda^n_{cd}q_d),
\\
J_{3\mu,R}^{(\mathbf 8)}&=&(\bar q_a\sigma_{\alpha\mu}\gamma_5\lambda^n_{ab}Q_b)(\bar{Q}_c\gamma^{\alpha}\lambda^n_{cd}q_d), \label{eq:88R}
\\ \nonumber
J_{4\mu,R}^{(\mathbf 8)}&=&-(\bar q_a\sigma_{\alpha\mu}\lambda^n_{ab}Q_b)(\bar{Q}_c\gamma^{\alpha}\gamma_5\lambda^n_{cd}q_d).
\end{eqnarray}
In these expressions $\lambda^n$ are eight Gell-Mann color matrices.

These 16 molecule fields with $J^P=1^+$ in Eqs.~\eqref{eq:11L}--\eqref{eq:88R}
are independent, but they do not have definite charge-conjugation parities.
We can use them to compose the molecular currents with definite charge-conjugation parities. The molecular currents with negative charge conjugation parity are
\begin{eqnarray}
J_{i\mu}^{(\mathbf 1)}=J_{i\mu,L}^{(\mathbf 1)}-J_{i\mu,R}^{(\mathbf 1)}, J_{i\mu}^{(\mathbf 8)}=J_{i\mu,L}^{(\mathbf 8)}-J_{i\mu,R}^{(\mathbf 8)}, i=1,\cdots,4, \label{eq:1+-}
\end{eqnarray}
and the molecular currents with positive charge conjugation parity are
\begin{eqnarray}
J_{i\mu}^{\prime (\mathbf 1)}=J_{i\mu,L}^{(\mathbf 1)}+J_{i\mu,R}^{(\mathbf 1)}, J_{i\mu}^{\prime (\mathbf 8)}=J_{i\mu,L}^{(\mathbf 8)}+J_{i\mu,R}^{(\mathbf 8)}, i=1,\cdots,4.
\end{eqnarray}

In this paper, we will study the $Z_c$ states by using the molecular currents constructed in Eq. \eqref{eq:1+-} with quantum numbers $J^{PC}=1^{+-}$
\begin{eqnarray}
\nonumber J_{1\mu}^{(\mathbf 1)}&=&(\bar q_a\gamma_5Q_a)(\bar{Q}_b\gamma_{\mu}q_b)+(\bar q_a\gamma_{\mu}Q_a)(\bar{Q}_b\gamma_5q_b),
\\ \nonumber J_{2\mu}^{(\mathbf 1)}&=&(\bar q_aQ_a)(\bar{Q}_b\gamma_{\mu}\gamma_5q_b)-(\bar q_a\gamma_{\mu}\gamma_5Q_a)(\bar{Q}_bq_b),
\\ \nonumber
J_{3\mu}^{(\mathbf 1)}&=&(\bar q_a\gamma^{\alpha}Q_a)(\bar{Q}_b\sigma_{\alpha\mu}\gamma_5q_b)-(\bar q_a\sigma_{\alpha\mu}\gamma_5Q_a)(\bar{Q}_b\gamma^{\alpha}q_b),
\\ \nonumber
J_{4\mu}^{(\mathbf 1)}&=&(\bar q_a\gamma^{\alpha}\gamma_5Q_a)(\bar{Q}_b\sigma_{\alpha\mu}q_b)+(\bar q_a\sigma_{\alpha\mu}Q_a)(\bar{Q}_b\gamma^{\alpha}\gamma_5q_b),
\\ J_{1\mu}^{(\mathbf 8)}&=&(\bar q_a\gamma_5\lambda^n_{ab} Q_b)(\bar{Q}_c\gamma_{\mu}\lambda^n_{cd}q_d)+(\bar q_a\gamma_{\mu}\lambda^n_{ab}Q_b)(\bar{Q}_c\gamma_5\lambda^n_{cd}q_d), \label{eq:currents1+-}
\\ \nonumber J_{2\mu}^{(\mathbf 8)}&=&(\bar q_a\lambda^n_{ab}Q_b)(\bar{Q}_c\gamma_{\mu}\gamma_5\lambda^n_{cd}q_d)-(\bar q_a\gamma_{\mu}\gamma_5\lambda^n_{ab}Q_b)(\bar{Q}_c\lambda^n_{cd}q_d),
\\ \nonumber
J_{3\mu}^{(\mathbf 8)}&=&(\bar q_a\gamma^{\alpha}\lambda^n_{ab}Q_b)(\bar{Q}_c\sigma_{\alpha\mu}\gamma_5\lambda^n_{cd}q_d)-(\bar q_a\sigma_{\alpha\mu}\gamma_5\lambda^n_{ab}Q_b)(\bar{Q}_c\gamma^{\alpha}\lambda^n_{cd}q_d),
\\ \nonumber
J_{4\mu}^{(\mathbf 8)}&=&(\bar q_a\gamma^{\alpha}\gamma_5\lambda^n_{ab}Q_b)(\bar{Q}_c\sigma_{\alpha\mu}\lambda^n_{cd}q_d)+(\bar q_a\sigma_{\alpha\mu}\lambda^n_{ab}Q_b)(\bar{Q}_c\gamma^{\alpha}\gamma_5\lambda^n_{cd}q_d),
\end{eqnarray}
in which $J_{1\mu}^{(\mathbf 1)}-J_{4\mu}^{(\mathbf 1)}$ belong to the color structure $\left(\mathbf{1}_{[q\bar Q]}\otimes\mathbf{1}_{[Q\bar q]}\right)$
while $J_{1\mu}^{(\mathbf 8)}-J_{4\mu}^{(\mathbf 8)}$ belong to the color structure $\left(\mathbf{8}_{[q\bar Q]}\otimes\mathbf{8}_{[Q\bar q]}\right)$.
The eight independent diquark-antidiquark tetraquark fields with $J^{PC}=1^{+-}$ constructed in Ref.~\cite{2011-Chen-p34010-34010} 
can
be written as combinations of these eight meson-meson type molecule currents. Moreover, one can construct other eight ``meson-meson'' type molecule currents,
having the color structure $[\bar q_a Q_b][\bar Q_b q_a]$. They can also be written as combinations of these eight meson-meson type molecule currents, having the color structures
$[\bar q_a Q_a][\bar Q_b q_b]$ and $[\bar q_a \lambda^n_{ab} Q_b][\bar Q_c \lambda^n_{cd} q_d]$.
In general, there is no one to one correspondence between the current and the state. The independence of the currents means that if the physical state is a molecular
state, it would be best to choose a molecular type of current so that it has a large overlap with the physical state. Similarly, it would be best to choose
a tetraquark current for a tetraquark state.

We note that the interpolating currents listed in Eq.~\eqref{eq:currents1+-} should contain the quark contents $\bar uc\bar cu+\bar dc\bar cd$ to be
neutral molecular currents of $I=1$. Such molecular currents have quantum numbers $I^GJ^{PC}=1^+1^{+-}$ and thus couple to neutral $Z_c$ states.
The corresponding operators with the quark contents $\bar uc\bar cd$ or
$\bar dc\bar cu$ can couple to charged $Z_c$ states. They altogether form isospin triplets. However, we will work in the $SU(2)$ isospin symmetry without considering
the effect of isospin breaking in this paper, i.e., we neglect instantons because we are in the vector channel, the masses of the up and down quarks and maintain isospin for the quark condensates  $\uu=\dd=\qq$,
$\uGua=\dGda=\qGqa$. Accordingly, the QCD sum rules for any iso-triplet are the same. Moreover, the isoscalar molecular currents can also be obtained from
Eq.~\eqref{eq:currents1+-} with the quark contents $\bar uc\bar cu-\bar dc\bar cd$, and the sum rules for theses currents are also
the same as those for the isospin triplet currents. Therefore, the same mass predictions would be obtained for the neutral and charged
$Z_c$ states with $I^GJ^{P(C)}=1^+1^{+(-)}$ and their isoscalar partner with $I^GJ^{PC}=0^-1^{+-}$. This expectation is reasonable for these quarkonium-like states,
for example, the neutral states $Z_c^0(3900)$ \cite{2013-Xiao-p366-370} and $Z_c^0(4020)$ \cite{2014-Ablikim-p212002-212002} lie very close to their charged partner
$Z_c(3900)^+$ and $Z_c(4020)^+$, respectively.

\section{QCD sum rules formalism}\label{Sec:QSR}
With these currents constructed in Eq.~\eqref{eq:currents1+-},
we can study the following two-point correlation function
\begin{eqnarray}
\nonumber\Pi_{\mu \nu}\left(q^2\right)&=& i \int d^4x e^{iq\cdot x} \langle 0 | T\left[J_{\mu}(x) J_{\nu}^\dagger (0)\right] | 0 \rangle
\label{def:pi}
\\ &=&\Pi\left(q^2\right)\left(\frac{q_\mu q_\nu}{q^2}-g_{\mu\nu}\right)+\Pi^\prime\left(q^2\right)\frac{q_{\mu}q_{\nu}}{q^2} \, ,
\end{eqnarray}
where $\Pi(q^2)$ and $\Pi^\prime(q^2)$ are invariant functions related to spin-1 and spin-0 states, respectively.
The two-point correlation function can be described
at both the hadron and quark-gluon levels. At the hadron level, the correlation function has a dispersion
relation representation
\begin{eqnarray}
\Pi(q^2)=\frac{(q^2)^N}{\pi}\int_{s<}^{\infty}\frac{{\rm Im}\Pi(s)}{s^N(s-q^2-i\epsilon)}ds+\sum_{n=0}^{N-1}b_n(q^2)^n\, , \label{Phenpi}
\end{eqnarray}
in which $b_n$ are the unknown subtraction constants which can be removed by taking the Borel transform.
The lower limit $s_<$ denotes a physical threshold. With this expression, one only needs to evaluate the imaginary
part of the correlation function, which is much easier than the full calculation.
The imaginary part of the correlation function is defined as the spectral function $\rho(s)=\frac{1}{\pi}{\rm Im}\Pi(s)$,
which is usually evaluated at the hadron level by inserting intermediate hadron states
\begin{eqnarray}
\nonumber
\rho(s)&\equiv&\sum_n\delta(s-m_n^2)\langle0|\eta|n\rangle\langle n|\eta^+|0\rangle\\
&=&f_X^2\delta(s-m_X^2)+ \mbox{continuum}\, ,   \label{Phenrho}
\end{eqnarray}
where we have adopted the usual pole plus continuum parametrization in the second step. All the intermediate states $|n\rangle$ must have
the same quantum numbers as the interpolating currents $J_{\mu}(x)$. The lowest-lying resonance with hadron mass $m_X$ couples to the current
$J_{\mu}(x)$ via
\begin{eqnarray}
\langle0|J_{\mu}|X\rangle=f_X\epsilon_{\mu}\, , \label{coupling_parameter}
\end{eqnarray}
in which $f_X$ is coupling constant and $\epsilon_{\mu}$ is the polarization vector ($\epsilon\cdot q=0$).

At the quark-gluon level, we evaluate the correlation function and spectral density via the QCD operator product expansion (OPE) up to
dimension-eight at the leading order of $\alpha_s$. The correlation function $\Pi(q^2)$ and spectral density $\rho(s)$ can be expressed in
terms of quark and gluon fields. These results are compared with Eq.~\eqref{Phenpi} obtained at the
hadron level to establish sum rules for hadron parameters, such as masses, magnetic moments and coupling constants of ground state hadrons.
As mentioned above, we usually take Borel transform to the correlation functions at both the hadron level and quark-gluon level to remove
the unknown constants in Eq.~\eqref{Phenpi} and suppress the continuum contributions. Using the spectral function defined in Eq.~\eqref{Phenrho},
the sum rules can be obtained as
\begin{eqnarray}
\mathcal{L}_{k}\left(s_0, M_B^2\right)=\int_{s<}^{s_0}dse^{-s/M_B^2}\rho(s)s^k=f_X^2m_X^{2k}e^{-m_X^2/M_B^2}\, , \label{sumrule}
\end{eqnarray}
where $M_B$ is the Borel parameter and $\rho(s)$ in the integral is the spectral density evaluated in QCD side. The upper integral limit
$s_0$ is the continuum threshold above which the contributions from the continuum and higher excited states can be approximated well by
the spectral function. Finally, the hadron mass $m_X$ for the lowest-lying state is extracted as
\begin{eqnarray}
m_X=\sqrt{\frac{\mathcal{L}_{1}\left(s_0, M_B^2\right)}{\mathcal{L}_{0}\left(s_0, M_B^2\right)}}\, . \label{mass}
\end{eqnarray}
It is shown in Eqs.~\eqref{sumrule} and \eqref{mass} that the extracted hadron mass $m_X$ is a function of the continuum threshold
$s_0$ and Borel mass $M_B$. One can perform the QCD sum rule analysis with these two equations. At the leading order in $\alpha_s$,
the spectral density $\rho(s)$ in Eq.~\eqref{sumrule} is evaluated up to dimension eight, including the perturbative term,
quark condensate $\qq$, gluon condensate $\GGb$, quark-gluon mixed condensate $\qGqb$, four-quark condensate $\qq^2$ and
the dimension eight condensate $\qq\qGqb$
\begin{eqnarray}
\rho(s)&=&\rho^{pert}(s)+\rho^{\qq}(s)+\rho^{\GGa}(s)+\rho^{\qq^2}(s)+\rho^{\qGqa}(s)+\rho^{\qq\qGqa}(s). \label{OPE}
\end{eqnarray}
To illustrate our numerical analysis, we use the current $J_{1\mu}^{(\mathbf 8)}$ as an example and show its spectral density in the following:
\begin{eqnarray}
\nonumber
\rho^{{(\mathbf 8)}pert}_1(s)&=&\frac{1}{192\pi^6}\dab\frac{(1-\alpha-\beta)(1+\alpha+\beta)\f(s)^4}{\alpha^3\beta^3}\, ,
\non
\rho^{{(\mathbf 8)}\qq}_1(s)&=&-\frac{m_Q\qq}{6\pi^4}\dab\frac{(1-\alpha-\beta)\f(s)\left[3m_Q^2(\alpha+\beta)-7\alpha\beta s\right]}{\alpha\beta^2}\, ,
\non
\rho^{{(\mathbf 8)}\GGa}_1(s)&=&-\frac{\GGb}{288\pi^6}\dab\Bigg\{\frac{7m_Q^2(1-\alpha-\beta)^2(5+\alpha+\beta)\f(s)}{96\alpha^2\beta^2}
\non&&+\frac{7m_Q^2(1+\alpha+\beta)\f(s)}{16\alpha\beta}-\frac{m_Q^2(1-\alpha-\beta)^2\left[m_Q^2(\alpha+\beta)-2\alpha\beta s\right]}{\alpha^3}
\non&&-\frac{(1-\alpha-\beta)\left[m_Q^2(3+\alpha+\beta)+2\alpha\beta s\right]\f(s)}{8\alpha^2\beta}\Bigg\}\, ,
\\
\rho^{{(\mathbf 8)}\qGqa}_1(s)&=&\frac{m_Q\qGqb}{144\pi^4}\dab
\non&& \Bigg\{\frac{(1-\alpha-\beta)\left[6m_Q^2(\alpha+\beta)-11\alpha\beta s\right]}{\beta^2}
+\frac{17\left[3m_Q^2(\alpha+\beta)-5\alpha\beta s\right]}{2\beta}\Bigg\}\, ,
\non
\rho^{{(\mathbf 8)}\qq^2}_1(s)&=&\frac{4m_Q^2\qq^2}{9\pi^2}\sqrt{1-4m_Q^2/s}\, ,
\non
\rho^{{(\mathbf 8)}\qq\qGqa}_1(s)&=&\frac{\qq\qGqb}{108\pi^2}\int_0^1d\alpha
\Bigg\{\frac{24m_Q^4}{\alpha^2}\delta'\left(s-\tilde{m}^2_Q\right)+
\frac{m_Q^2(5\alpha^2-6\alpha+3)}{\alpha(1-\alpha)^2}\delta\left(s-\tilde{m}^2_Q\right)-\alpha H\left(s-\tilde{m}^2_Q\right)
\Bigg\}\, , \label{spectral density}
\end{eqnarray}
in which $\tilde{m}^2_Q=\frac{m_Q^2}{\alpha(1-\alpha)}$, $\delta'\left(s-\tilde{m}^2_Q\right)=\frac{d\delta\left(s-\tilde{m}^2_Q\right)}{ds}$,
and $H(s-\tilde{m}^2_Q)$ is a Heaviside step function. The integration limits are $\alpha_{min}=\frac{1-\sqrt{1-4m_Q^2/s}}{2}$,
$\beta_{min}=\frac{\alpha m_Q^2}{\alpha s-m_Q^2}$, $\beta_{max}=1-\alpha$, $\alpha_{max}=\frac{1+\sqrt{1-4m_Q^2/s}}{2}$ and $m_Q$ is
the heavy quark mass. We note that we have ignored the chirally suppressed terms with the light quark mass and adopted the factorization assumption
of vacuum saturation for higher dimensional condensates ($D=6$ and $D=8$). The results for other currents listed in Eq.~\eqref{eq:currents1+-} are collected in Appendix.~\ref{sec:rhos}.
From these expressions we can find that the $D=3$ quark condensate $\qq$ and the $D=5$ mixed condensate $\qGqb$ are both multiplied by the heavy quark mass $m_Q$, which are thus
important power corrections to the correlation functions.

\section{Numerical Analysis}\label{sec:NA}

In this section we still use the current $J_{1\mu}^{(\mathbf 8)}$ as an example and perform the numerical analysis.
The following QCD parameters of quark masses and various condensates are used in our
analysis~\cite{2014-Olive-p90001-90001,2001-Eidemuller-p203-210, 1999-Jamin-p300-303,2002-Jamin-p237-243,2011-Khodjamirian-p94031-94031}:
\begin{eqnarray}
\nonumber
&&m_c(m_c)=(1.23\pm0.09)\text{ GeV} \, , \non
&&m_b(m_b)=(4.20\pm0.07)\text{ GeV} \, , \non
&&\qq=-(0.23\pm0.03)^3\text{ GeV}^3 \, ,
\\ &&\qGqb=-M_0^2\qq\, ,
\non &&M_0^2=(0.8\pm0.2)\text{ GeV}^2 \, ,
\non&&\GGb=(0.88\pm0.14)\text{
GeV}^4 \, . \label{parameters}
\end{eqnarray}
Note that there is a minus sign implicitly included in the definition of the coupling constant $g_s$. We use the running masses in the $\overline{MS}$
scheme for the charm and bottom quarks.

As mentioned above, the extracted hadron mass $m_X$ in Eq.~\eqref{mass} is a function of the continuum threshold $s_0$ and the Borel mass $M_B$, which
are two vital parameters in QCD sum rule analyses. If the final result, $m_X$, does
not depend on these two free parameters, then the method of QCD sum rules would have perfect predictive power. However, reliable mass predictions are
obtained when there is weak dependance on these parameters in a reasonable working regions. Principally, there are two criteria to find a Borel window(reasonable working
region of $M_B$): the requirement of the OPE convergence results in a lower bound while the constraint of the pole contribution
leads to an upper bound. At the same time, we will study the variation of the hadron mass $m_X$ with respect to the continuum threshold.
An optimized value of the continuum threshold $s_0$ is chosen to minimize the dependence of the extracted hadron mass $m_X$ on the Borel mass $M_B$.

In Eq.~\eqref{OPE}, the non-perturbative terms are evaluated up to dimension eight. After the numerical analysis, we find that
the quark condensate $\qq$ and quark--gluon mixed condensate $\qGqb$ are dominant power corrections while the contributions of
other condensates are much smaller. Using the spectral density for the current $J_{1\mu}^{(\mathbf 8)}$ in
Eq.~\eqref{spectral density},
we show the contribution of the dimension eight condensate to the correlation function $\Pi_1^{\qq\qGqa}(M_B^2, s_0)/\Pi_1^{all}(M_B^2, s_0)$
in Fig.~\ref{figOPE1} with $s_0\to\infty$, in which the ratio decreases with respect to $M_B^2$. Accordingly, we require the dimension eight
condensate contribution to be less than $5\%$, which results in a lower bound $M_{min}^2=2.8$ GeV$^2$.
\begin{center}
\includegraphics[scale=0.7]{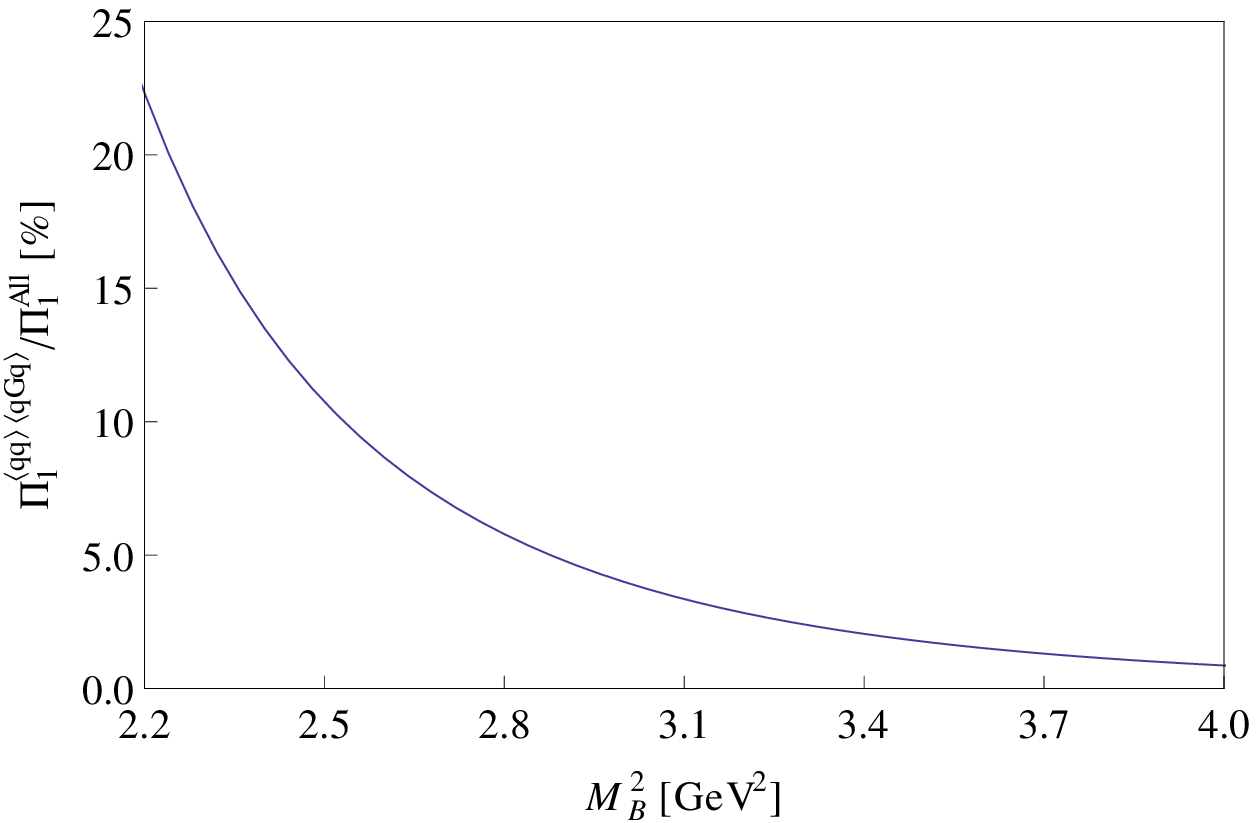}
\figcaption{OPE convergence for the current $J_{1\mu}^{(\mathbf 8)}$ with $s_0\to\infty$.} \label{figOPE1}
\end{center}

To determine the upper bound on $M_B^2$,
we define the pole contribution(PC) using the sum rules established in Eq. \eqref{sumrule},
\begin{eqnarray}
\text{PC}(s_0, M_B^2)=\frac{\mathcal{L}_{0}\left(s_0, M_B^2\right)}{\mathcal{L}_{0}\left(\infty, M_B^2\right)}, \label{PC}
\end{eqnarray}
which represents the lowest-lying resonance contribution to the correlation function. The continuum threshold $s_0$ is
an important parameter to the pole contribution. We study the variation of $m_X$ with $s_0$ in the left panel of Fig.~\ref{fig1_8cc} by
varying the value of $M_B^2$ from its lower bound. With different values of $M_B^2$, there curves intersect at $s_0=18$ GeV$^2$ around which
the variation of $m_X$ with $M_B^2$ reaches it minimum. This is thus an optimized value of the continuum threshold $s_0$ to study the pole contribution
defined in Eq.~\eqref{PC}. Requiring that PC be larger than $20\%$, we obtain an upper bound on the Borel mass $M_{max}^2=3.7\,{\rm GeV^2}$.
The Borel window is then determined to be $2.8\,{\rm GeV^2}\leq M_B^2\leq3.7\,{\rm GeV^2}$ with the threshold value $s_0=18$ GeV$^2$.

In the right panel of Fig.~\ref{fig1_8cc}, we show the variation of the hadron mass $m_X$ with respect to $M_B^2$. The Borel window varies
quickly for different value of $s_0$. One notes that the mass curves decrease significantly in the region $M_B^2\leq M_{min}^2$ while becoming quite stable
inside the Borel windows. Finally, we can extract the hadron mass and the coupling constant
\begin{eqnarray}
m_{X_{1}^{(\mathbf 8)}}&=&(3.90\pm 0.12)~\text{GeV}\, ,
\\
f_{X_{1}^{(\mathbf 8)}}&=&(0.69\pm 0.21)\times 10^{-2}~\text{GeV}^5 \, .
\end{eqnarray}
This value is consistent with the mass of $Z_c(3900)$, which implies the possible molecule interpretation of this
new resonance. Here we would like to emphasize that in our calculations we have not used masses of heavy-light mesons as inputs, but simply note
that the obtained value 3.9 GeV (as well as other $m_X$ listed in Table~\ref{table1}) is close to the threshold of two heavy-light mesons. Further studies are needed to understand
whether there is an underlying reason for these results.

\begin{center}
\begin{tabular}{lr}
\scalebox{0.6}{\includegraphics{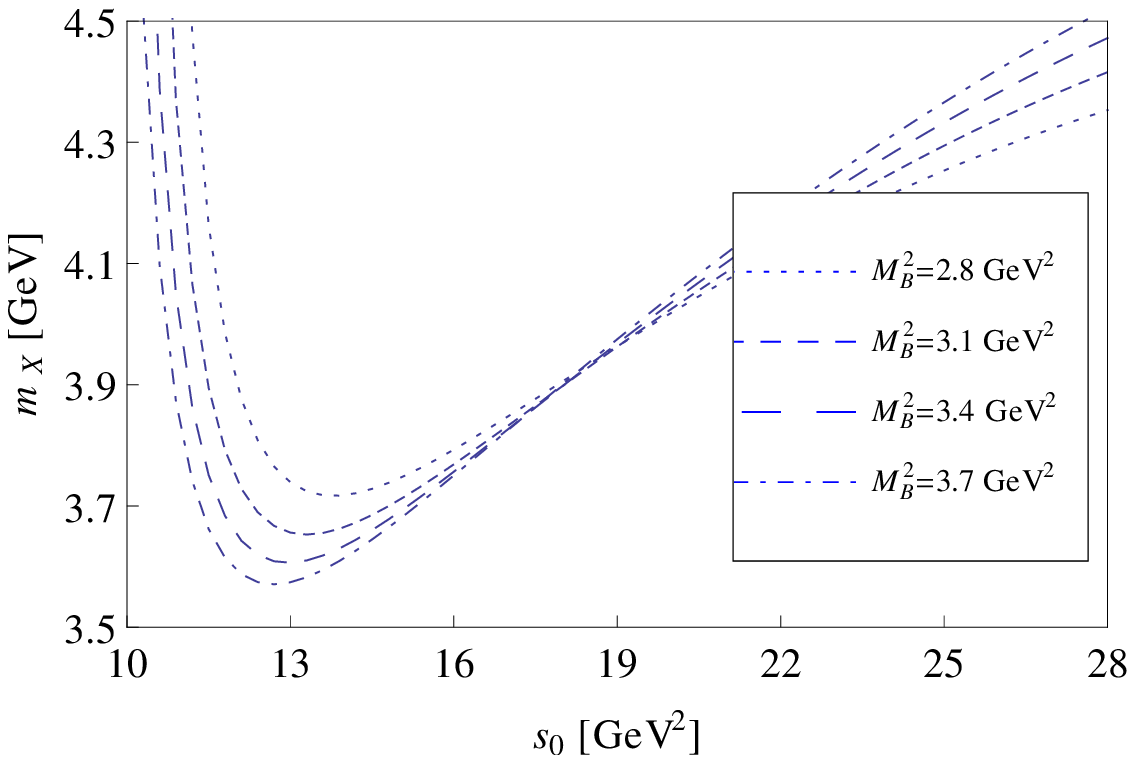}}&
\scalebox{0.6}{\includegraphics{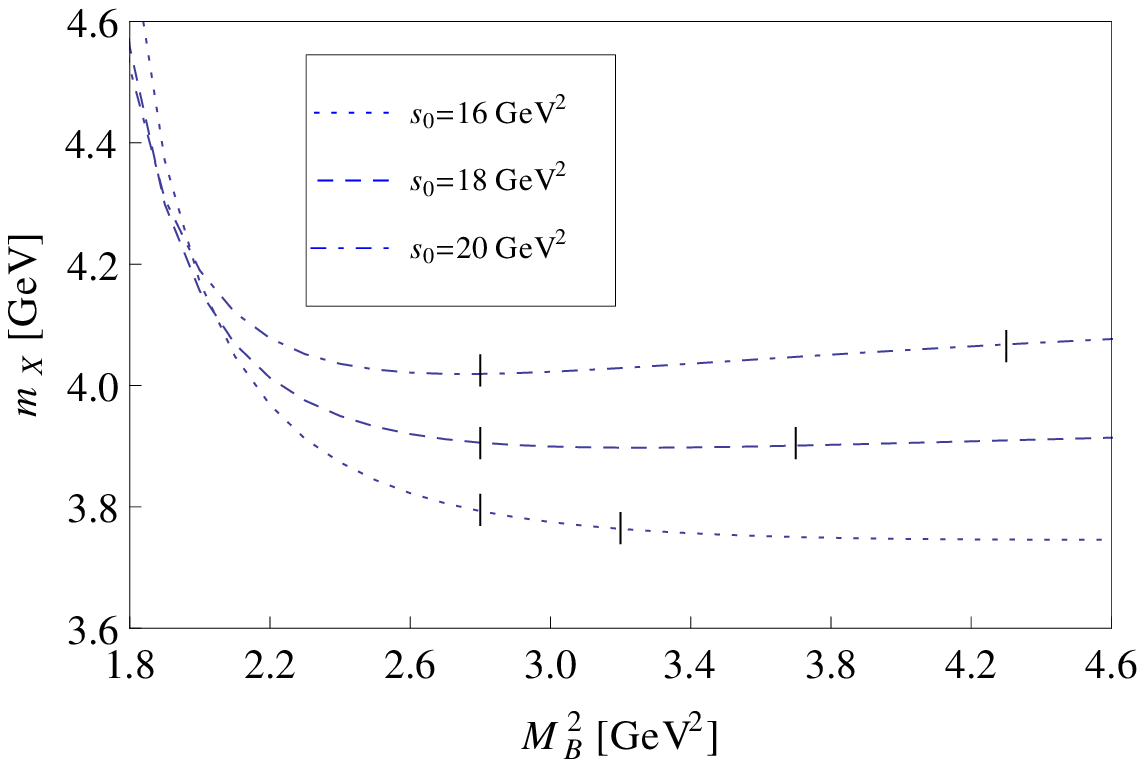}}
\end{tabular}
\figcaption{Variations of the charmonium-like molecule hadron mass $m_X$ with $s_0$ and $M_B^2$ for the current $J_{1\mu}^{(\mathbf 8)}$.} \label{fig1_8cc}
\end{center}

After performing similar numerical analyses for the other interpolating currents, we collect the extracted numerical results for the hadron masses and
coupling constants in Table \ref{table1}. The mass sum rules for the currents $J_{2\mu}^{(\mathbf 1)}$ and $J_{2\mu}^{(\mathbf 8)}$ are unstable
and thus they do not give reliable mass predictions. For the current $J_{1\mu}^{(\mathbf 1)}$, the lower bound of the Borel window is very small
under the first (convergence) criterion. Although it leads to very good OPE convergence and broad Borel window, we
need to consider the stability of the Borel curves, from which the lower bound of the Borel window is determined to be $3.3$ GeV$^2$. The situations
for the currents $J_{4\mu}^{(\mathbf 1)}$ and $J_{4\mu}^{(\mathbf 8)}$ are very different, in which the lower bounds on $M_B^2$ are bigger than their
upper bounds, suggesting the OPE convergence is poor for them. By loosening the criterion of the OPE convergence and requiring the dimension eight contribution to be less than $10\%$,
we can still obtain stable mass sum rules for $J_{4\mu}^{(\mathbf 1)}$ and $J_{4\mu}^{(\mathbf 8)}$ and reliable mass predictions, as shown in Table~\ref{table1}.
The error sources including the uncertainties of the various parameters in Eq.~\eqref{parameters} and the continuum threshold $s_0$ are considered
to obtain the errors for hadron masses and coupling constants.
\begin{center}
\begin{tabular}{cccccc}
\hlinewd{.8pt}
& Current ~~&~~ $s_0(\mbox{GeV}^2)$ ~~&~~ \mbox{Borel window} $(\mbox{GeV}^2)$ ~~&~~ $m_X$ \mbox{(GeV)} ~~&~~ $f_X$ $(10^{-2}\mbox{GeV}^5)$ \\
\hline
& $J_{1\mu}^{(\mathbf 1)}$   & 21   & $3.3 - 4.5 $ & $4.22\pm0.14$ & $0.57\pm0.16$ \\
& $J_{2\mu}^{(\mathbf 1)}$   & $-$  & $    -     $ & $     -     $ & $   -   $ \\
& $J_{3\mu}^{(\mathbf 1)}$   & 20   & $3.3 - 4.2 $ & $4.04\pm0.12$ & $0.93\pm0.27$  \\
& $J_{4\mu}^{(\mathbf 1)}$   & 20   &$(3.0 - 3.3)$ & $4.02\pm0.15$ & $0.35\pm0.13$ \\
\vspace{5pt}\\
& $J_{1\mu}^{(\mathbf 8)}$   & 18   & $2.8 - 3.7 $ & $3.90\pm0.12$ & $0.69\pm0.21$ \\
& $J_{2\mu}^{(\mathbf 8)}$   & $-$  & $    -     $ & $     -     $ & $   -   $ \\
& $J_{3\mu}^{(\mathbf 8)}$   & 18   & $3.1 - 3.9 $ & $3.85\pm0.11$ & $1.51\pm0.46$ \\
& $J_{4\mu}^{(\mathbf 8)}$   & 20   &$(2.8 - 3.1)$ & $4.03\pm0.18$ & $0.59\pm0.23$ \\
\hline
\hlinewd{.8pt}
\end{tabular}
\tabcaption{Numerical results for the charmonium-like molecule states. \label{table1}}
\end{center}

In Table \ref{table1}, the extracted masses from the currents of color structure $\left(\mathbf{8}_{[q\bar Q]}\otimes\mathbf{8}_{[Q\bar q]}\right)$
are about $3.85 - 4.03$ GeV, which are slightly below the $4.02 - 4.22$ GeV from the currents of color structure $\left(\mathbf{1}_{[q\bar Q]}\otimes\mathbf{1}_{[Q\bar q]}\right)$,
although they both lie precisely in the range of spectra of $Z_c$ states. The masses extracted from the
currents $J_{1\mu}^{(\mathbf 8)}$ and $J_{2\mu}^{(\mathbf 8)}$ are $(3.90\pm0.12)$ GeV and $(3.85\pm0.11)$ GeV respectively, which are
clearly consistent with the mass of $Z_c(3900)$. The interpolating currents $J_{3\mu}^{(\mathbf 1)}$, $J_{4\mu}^{(\mathbf 1)}$ and
$J_{4\mu}^{(\mathbf 8)}$ give hadron masses $(4.04\pm0.12)$ GeV, $(4.02\pm0.15)$ GeV and $(4.03\pm0.18)$ GeV respectively, which
are in very close proximity to the masses of the $Z_c(4020)$ and $Z_c(4050)$ mesons, although the latter state is not confirmed to date.
We note that these values are also in rough agreement with the mass of $Z_c(3900)$ state.
However, one can find that it is better to chose the currents $J_{1\mu}^{(\mathbf 8)}$ and $J_{2\mu}^{(\mathbf 8)}$ to fit the mass of $Z_c(3900)$ because these two currents have a larger overlap with the physical state. We can infer that $Z_c(3900)$ has a structure well represented by the currents $J_{1\mu}^{(\mathbf 8)}$ and $J_{2\mu}^{(\mathbf 8)}$.
 Last but not least, the current $J_{1\mu}^{(\mathbf 1)}$ leads to a mass
prediction $(4.22\pm0.14)$ GeV, which is in good agreement with the mass of $Z_c(4200)$. The interpolating currents $J_{1\mu}^{(\mathbf 1)}$,
$J_{2\mu}^{(\mathbf 1)}$, $J_{3\mu}^{(\mathbf 1)}$ and $J_{4\mu}^{(\mathbf 1)}$ are constructed as $D\bar D^*$, $D_0^{\ast}\bar D_1$,
$D^*\bar D^*$ and $D_1\bar D_1$ molecular operators, respectively. Our results in Table \ref{table1} suggest a possible landscape of
hadronic molecular interpretations of the charged and neutral $Z_c$ states. According to the above analysis, we suggest
that the $Z_c(4200)$ meson to be a $D\bar D^*$ state, while the $Z_c(4020)$, $Z_c(4050)$ and $Z_c(3900)$ mesons to be $D^*\bar D^*$ or
$D_1\bar D_1$ states. The analysis of the current $J_{2\mu}^{(\mathbf 1)}$ may imply that the stable $D_0^{\ast}\bar D_1$ molecular state does not exist.

\begin{center}
\begin{tabular}{lr}
\scalebox{0.6}{\includegraphics{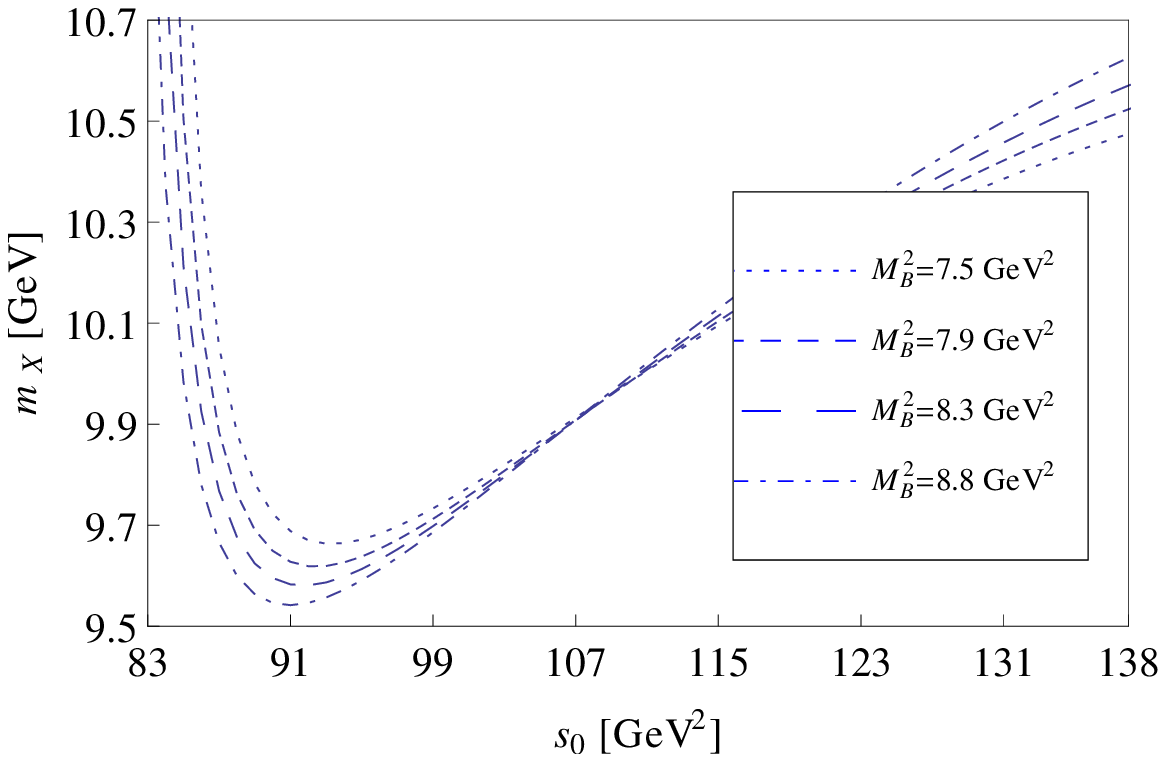}}&
\scalebox{0.6}{\includegraphics{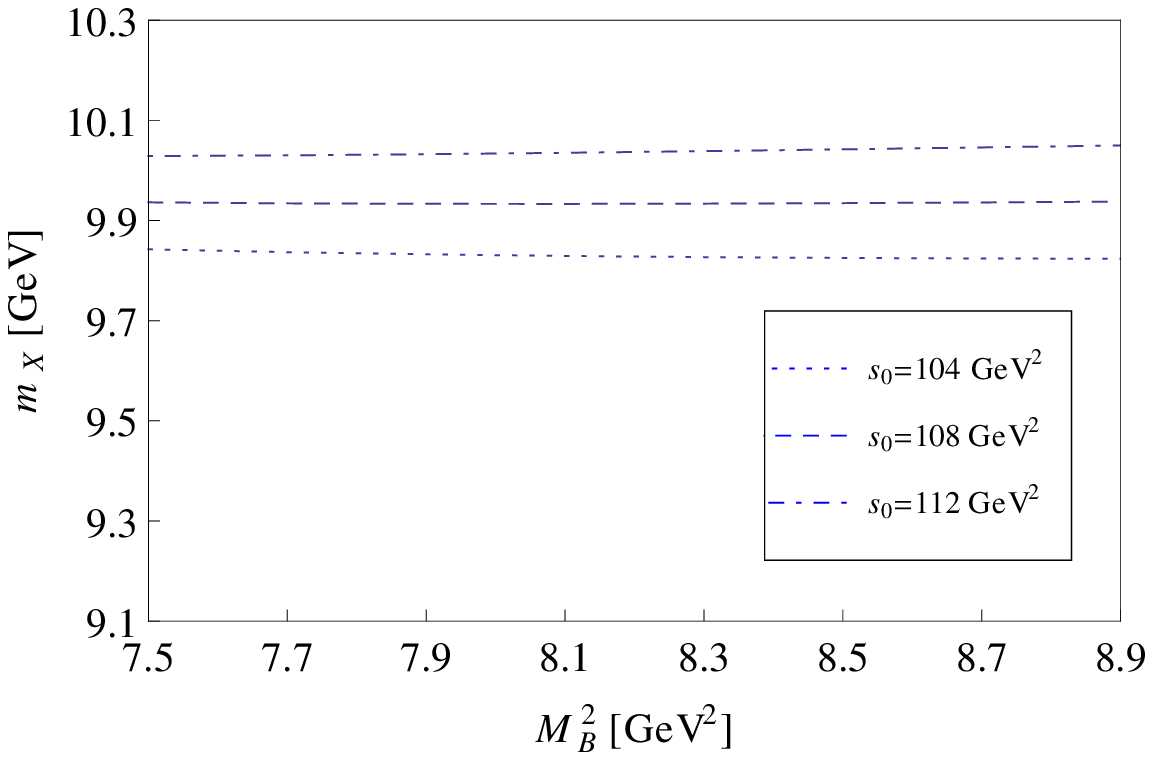}}
\end{tabular}
\figcaption{Variations of the bottomonium-like molecule hadron mass $m_X$ with $s_0$ and $M_B^2$ for the current $J_{1\mu}^{(\mathbf 8)}$.} \label{fig1_8bb}
\end{center}

Similarly, we can study bottomonium-like molecule states with $J^{PC}=1^{+-}$ by taking $m_Q=m_b$ in the expression of the spectral density in
Eq.~\eqref{spectral density} and Appendix~\ref{sec:rhos}. The bottomonium-like molecule system is similar to the charmonium-like system due to heavy quark symmetry.
Under the same criteria, one finds that the Borel window for the $q\bar bb\bar q$ system is much broader than the $q\bar cc\bar q$ system.
This suggests a stricter limitation of the pole contribution, which is only required to be larger than $20\%$ for the $q\bar cc\bar q$ systems.
For the $q\bar bb\bar q$ system with $J_{1\mu}^{(\mathbf 8)}$, we require the same OPE convergence criterion as the $q\bar cc\bar q$
system while modifying the requirement of the pole contribution to be larger than $30\%$. The Borel window is obtained as $7.5\,{\rm GeV^2}\leq
M_B^2\leq8.8\,{\rm GeV^2}$ with the continuum threshold $s_0=108$ GeV$^2$. Using these values of the parameters, we show the variations of the
hadron mass with respect to the Borel mass $M_B^2$ and the threshold value $s_0$ in Fig.~\ref{fig1_8bb}. The Borel curves are shown to be very stable and give reliable predictions
of the hadron mass and coupling constant
\begin{eqnarray}
m_{X_{1b}^{(\mathbf 8)}}&=&(9.93\pm 0.15)~\text{GeV}\, ,
\\
f_{X_{1b}^{(\mathbf 8)}}&=&(1.02\pm 0.30)\times 10^{-3}~\text{GeV}^5 \, .
\end{eqnarray}
After numerical analyses of all interpolating currents, we collect the numerical results for the $q\bar bb\bar q$ states in Table~\ref{table2}.
Similar to the charmonium-like system, there is no significant Borel window for the current $J_{4\mu}^{(\mathbf 8)}$ under the above criteria.
The Borel window $(7.5 - 8.5)$ GeV$^2$ written in parenthesis is obtained by loosening the requirement of the OPE convergence to be 10\%. However, the mass
prediction under this Borel window is still reliable. The extracted mass for the current $J_{1\mu}^{(\mathbf 1)}$ is about $10.48\pm0.15$ GeV,
which is consistent with the mass of the $Z_b(10610)$ meson within the error, supporting the $B\bar B^*$ molecule interpretation for this state.

\begin{center}
\begin{tabular}{cccccc}
\hlinewd{.8pt}
& Current ~~&~~ $s_0(\mbox{GeV}^2)$ ~~&~~ \mbox{Borel window} $(\mbox{GeV}^2)$ ~~&~~ $m_X$ \mbox{(GeV)} ~~&~~ $f_X$ $(10^{-3}\mbox{GeV}^5)$\\
\hline
& $J_{1\mu}^{(\mathbf 1)}$   & 121   & $8.0 - 11.0 $   & $10.48\pm0.15$   & $1.45\pm0.19$ \\
& $J_{2\mu}^{(\mathbf 1)}$   & $-$   & $    -      $   & $     -      $   & $     -     $ \\
& $J_{3\mu}^{(\mathbf 1)}$   & 113   & $8.0 - 9.4  $   & $10.14\pm0.15$   & $1.69\pm0.46$ \\
& $J_{4\mu}^{(\mathbf 1)}$   & 117   & $7.5 - 8.3  $   & $10.33\pm0.14$   & $0.58\pm0.34$ \\
\vspace{5pt}\\
& $J_{1\mu}^{(\mathbf 8)}$   & 108   & $7.5 - 8.8 $    & $ 9.93\pm0.15$   & $1.02\pm0.30$ \\
& $J_{2\mu}^{(\mathbf 8)}$   & $-$   & $    -     $    & $      -     $   & $     -     $ \\
& $J_{3\mu}^{(\mathbf 8)}$   & 108   & $7.8 - 8.7 $    & $ 9.92\pm0.15$   & $2.17\pm0.62$ \\
& $J_{4\mu}^{(\mathbf 8)}$   & 119   &$(7.5 - 8.5)$    & $10.46\pm0.14$   & $1.67\pm0.55$ \\
\hline
\hlinewd{.8pt}
\end{tabular}
\tabcaption{Numerical results for the bottomonium-like molecule states. \label{table2}}
\end{center}

\section{SUMMARY AND DISCUSSIONS}\label{sec:SUMMARY}

To study the charged exotic $Z_c$ and $Z_b$ states, we constructed all the charmonium-like/bottomonium-like molecular interpolating currents
with $J^{PC}=1^{+-}$, including both the singlet-singlet and octet-octet types of color structures. We calculated the two-point correlation
functions and spectral densities for these $q\bar QQ\bar q$ operators. Within the $SU(2)$ isospin symmetry, all the numerical results
of hadron masses and coupling constants in Tables \ref{table1} and \ref{table2} are suitable for the neutral and charged $Z_c$ states
with $I^GJ^{P(C)}=1^+1^{+(-)}$ and their isoscalar partner with $I^GJ^{PC}=0^-1^{+-}$.

At the leading order in $\alpha_s$, we calculated the two-point correlation functions and spectral densities up to dimension eight,
including the perturbative term, the quark condensate $\qq$, the mixed condensate $\qGqb$, the gluon condensate $\GGb$, the four-quark
condensate $\qq^2$ and the $D=8$ condensate $\qq\qGqb$. Being proportional to the heavy quark mass, the quark condensate $\qq$ is the
dominant power correction to the correlation function while the mixed condensate $\qGqb$ also gives an important contribution. After
performing the numerical analyses, we obtain reliable mass predictions $4.02 - 4.22$ GeV for the color singlet-singlet charmonium-like
molecular states and $3.85 - 4.03$ GeV for the color octet-octet ones. This mass spectrum of the $q\bar cc\bar q$
states is precisely consistent with the masses of the $Z_c$ states, suggesting a possible landscape of hadronic moleculear interpretations
of the newly observed $Z_c$ states. We suggest that the $Z_c(4200)$ meson is a $D\bar D^*$ state while the $Z_c(4020)$, $Z_c(4050)$
and $Z_c(3900)$ mesons is either a $D^*\bar D^*$ or $D_1\bar D_1$ state. The stable $D_0^{\ast}\bar D_1$ molecular state does not occur in our result.

The bottomonium-like $q\bar bb\bar q$ molecular states are also studied and the numerical results are collected in Table \ref{table2}.
The extracted masses are predicted to be around $9.92 - 10.48$ GeV, which are slightly lower than the masses of the charged $Z_b(10610)$
meson. However, the charged $Z_b(10610)$ meson is consistent with a $B\bar B^*$ molecular state within the theoretical uncertainties.

One finds that the hadron masses extracted from the color singlet-singlet currents are a bit higher than those extracted from the color
octet-octet currents, for both the charmonium-like and bottomonium-like systems. This situation is different from the result in
Ref. \cite{2014-Wang-p2891-2891}, in which the color octet-octet tetraquarks were heavier. In Refs. \cite{1996-Cho-p150-162,1996-Braaten-p730-733},
the color-octet mechanism was found to give contributions to quarkonia production via the emission or absorption of a soft gluon in NRQCD.
Similar mechanisms can be expected in the octet-octet quarkonium-like molecular systems, in which a color-octet $Q\bar q$ pair combines with
another color-octet $q\bar Q$ pair by exchanging a gluon.

The possible hadronic decay patterns of the $q\bar cc\bar q$ and $q\bar bb\bar q$ molecular states can be discussed by considering the
kinematic constraints and the conversations of parity, C-parity, isospin and G-parity.
Considering the hadron masses obtained in Table \ref{table1}, the possible S-wave
two-meson hadronic decay channels for the charmonium-like $q\bar cc\bar q$ molecular states with $I^GJ^{P(C)}=1^+1^{+(-)}$ are
\begin{eqnarray}
Z_c\to D\bar D^*, D^*\bar D^*, \eta_c(1S)\rho, J/\psi\pi, \psi(2S)\pi\, ,
\end{eqnarray}
and the possible P-wave decay channels are
\begin{eqnarray}
Z_c\to D^*_0\bar D, \eta_c(1S)b_1((1235)\, .
\end{eqnarray}
For their isoscalar $q\bar cc\bar q$ partners with $I^GJ^{PC}=0^-1^{+-}$, the possible S-wave decay channels are
\begin{eqnarray}
Z_c\to D\bar D^*, D^*\bar D^*, \eta_c(1S)\omega, J/\psi\eta, J/\psi\eta^{\prime}\, ,
\end{eqnarray}
while the P-wave decay channels are
\begin{eqnarray}
Z_c\to \eta_c(1S)h_1(1170), J/\psi f_0(980), J/\psi a_0(980), h_c(1P)\eta\, .
\end{eqnarray}

For the bottomonium-like $q\bar bb\bar q$ molecular states, the extracted masses in Table \ref{table2} lie below the open-bottom thresholds
so that only the hidden-flavor decay channels are kinematically allowed. The possible S-wave decay patterns for the $q\bar bb\bar q$ states
with $I^GJ^{P(C)}=1^+1^{+(-)}$ are $Z_b\to \Upsilon(1S)\pi, \Upsilon(2S)\pi$ while the S-wave decays are forbidden.
For the isoscalar partners with $I^GJ^{PC}=0^-1^{+-}$, their possible S-wave and P-wave decay channels are $Z_b\to \Upsilon(1S)\eta$ and
$Z_b\to \Upsilon(1S)f_0(980), \Upsilon(1S)a_0(980)$, respectively.

\section*{Acknowledgments}
This project is supported by the Natural Sciences and Engineering
Research Council of Canada (NSERC). H.X.C and S.L.Z. are supported
by the National Natural Science Foundation of China under Grants
No. 11205011, No. 11475015, and No. 11261130311.


\appendix

\section{Expressions of spectral density for other interpolating currents}\label{sec:rhos}
In Eq. \eqref{spectral density}, we have given the spectral density extracted from the current
$J_{1\mu}^{(\mathbf 8)}$. For other interpolating currents listed in Eq. \eqref{eq:currents1+-}, we
collect the expressions of the spectral density in this appendix up to dimension eight condensate,
as shown in \eqref{OPE}.
\begin{itemize}
\item For the current $J_{1\mu}^{(\mathbf 1)}$
{\allowdisplaybreaks
\begin{eqnarray}
\nonumber
\rho^{{(\mathbf 1)}pert}_1(s)&=&\frac{3}{2048\pi^6}\dab\frac{(1-\alpha-\beta)(1+\alpha+\beta)\f(s)^4}{\alpha^3\beta^3}\, ,
\non
\rho^{{(\mathbf 1)}\qq}_1(s)&=&-\frac{3m_Q\qq}{64\pi^4}\dab\frac{(1-\alpha-\beta)\f(s)\left[3m_Q^2(\alpha+\beta)-7\alpha\beta s\right]}{\alpha\beta^2}\, ,
\non
\rho^{{(\mathbf 1)}\GGa}_1(s)&=&-\frac{\GGb}{512\pi^6}\dab
\non&&\Bigg\{\frac{(1-\alpha-\beta)\f(s)s}{\alpha}
-\frac{m_Q^2(1-\alpha-\beta)^2\left[m_Q^2(\alpha+\beta)-2\alpha\beta s\right]}{2\alpha^3}\Bigg\}\, ,
\non
\rho^{{(\mathbf 1)}\qGqa}_1(s)&=&-\frac{3m_Q\qGqb}{64\pi^4}\dab
\non&&\Bigg\{\frac{(1-\alpha-\beta)\left[m_Q^2(\alpha+\beta)-2\alpha\beta s\right]}{\beta^2}-
\frac{7m_Q^2(\alpha+\beta)-11\alpha\beta s}{2\alpha\beta}\Bigg\}\, ,
\non
\rho^{{(\mathbf 1)}\qq^2}_1(s)&=&\frac{m_Q^2\qq^2}{8\pi^2}\sqrt{1-4m_Q^2/s}\, ,
\\
\rho^{{(\mathbf 1)}\qq\qGqa}_1(s)&=&\frac{\qq\qGqb}{16\pi^2}\int_0^1d\alpha
\Bigg\{\frac{m_Q^4}{\alpha^2}\delta'\left(s-\tilde{m}^2_Q\right)-
\frac{m_Q^2}{\alpha}\delta\left(s-\tilde{m}^2_Q\right)
\Bigg\}\, . \label{SD11}
\end{eqnarray}
}

\item For the current $J_{2\mu}^{(\mathbf 1)}$
{\allowdisplaybreaks
\begin{eqnarray}
\nonumber
\rho^{{(\mathbf 1)}pert}_2(s)&=&\frac{3}{2048\pi^6}\dab\frac{(1-\alpha-\beta)(1+\alpha+\beta)\f(s)^4}{\alpha^3\beta^3}\, ,
\non
\rho^{{(\mathbf 1)}\qq}_2(s)&=&\frac{3m_Q\qq}{64\pi^4}\dab\frac{(1-\alpha-\beta)\f(s)\left[3m_Q^2(\alpha+\beta)-7\alpha\beta s\right]}{\alpha\beta^2}\, ,
\non
\rho^{{(\mathbf 1)}\GGa}_2(s)&=&-\frac{\GGb}{512\pi^6}\dab
\non&& \Bigg\{\frac{(1-\alpha-\beta)\f(s)s}{\alpha}
-\frac{m_Q^2(1-\alpha-\beta)^2\left[m_Q^2(\alpha+\beta)-2\alpha\beta s\right]}{2\alpha^3}\Bigg\}\, ,
\non
\rho^{{(\mathbf 1)}\qGqa}_2(s)&=&\frac{3m_Q\qGqb}{64\pi^4}\dab
\non&&\Bigg\{\frac{(1-\alpha-\beta)\left[m_Q^2(\alpha+\beta)-2\alpha\beta s\right]}{\beta^2}-
\frac{7m_Q^2(\alpha+\beta)-11\alpha\beta s}{2\alpha\beta}\Bigg\}\, ,
\non
\rho^{{(\mathbf 1)}\qq^2}_2(s)&=&\frac{m_Q^2\qq^2}{8\pi^2}\sqrt{1-4m_Q^2/s}\, ,
\\
\rho^{{(\mathbf 1)}\qq\qGqa}_2(s)&=&\frac{\qq\qGqb}{16\pi^2}\int_0^1d\alpha
\Bigg\{\frac{m_Q^4}{\alpha^2}\delta'\left(s-\tilde{m}^2_Q\right)-
\frac{m_Q^2}{\alpha}\delta\left(s-\tilde{m}^2_Q\right)
\Bigg\}\, . \label{SD12}
\end{eqnarray}
}
\item For the current $J_{3\mu}^{(\mathbf 1)}$
{\allowdisplaybreaks
\begin{eqnarray}
\nonumber
\rho^{{(\mathbf 1)}pert}_3(s)&=&\frac{1}{512\pi^6}\dab\Bigg\{\frac{9(1-\alpha-\beta)(1+\alpha+\beta)\f(s)^4}{4\alpha^3\beta^3}
\non&&-\frac{m_Q^2(1-\alpha-\beta)^2(5+\alpha+\beta)\f(s)^3}{\alpha^3\beta^3}\Bigg\}\, ,
\non
\rho^{{(\mathbf 1)}\qq}_3(s)&=&-\frac{9m_Q\qq}{64\pi^4}\dab\frac{(1-\alpha-\beta)\f(s)\left[3m_Q^2(\alpha+\beta)-7\alpha\beta s\right]}{\alpha\beta^2}\, ,
\non
\rho^{{(\mathbf 1)}\GGa}_3(s)&=&\frac{\GGb}{1024\pi^6}\dab\Bigg\{\frac{3m_Q^2(1-\alpha-\beta)^2\left[m_Q^2(\alpha+\beta)-2\alpha\beta s\right]}{\alpha^3}
\non&&-\frac{m_Q^2(1-\alpha-\beta)^2(5+\alpha+\beta)\left\{3(\alpha^2+\beta^2)\f(s)+m_Q^2(\alpha^3+\beta^3)\right\}}{6\alpha^3\beta^3}
\non&&+\frac{(1-\alpha-\beta)\f(s)\left[m_Q^2(3+\alpha+\beta)+2\alpha\beta s\right]}{\alpha^2\beta}\Bigg\}\, ,
\non
\rho^{{(\mathbf 1)}\qGqa}_3(s)&=&\frac{m_Q\qGqb}{64\pi^4}\dab
\non&&\Bigg\{\frac{(1-\alpha-\beta)\left[3m_Q^2(\alpha+\beta)-4\alpha\beta s\right]}{\beta^2}
+\frac{(2+7\alpha-2\beta)\left[3m_Q^2(\alpha+\beta)-5\alpha\beta s\right]}{2\alpha\beta}\Bigg\}\, ,
\non
\rho^{{(\mathbf 1)}\qq^2}_3(s)&=&\frac{5(2m_Q^2+s)\qq^2}{48\pi^2}\sqrt{1-4m_Q^2/s}\, ,
\non
\rho^{{(\mathbf 1)}\qq\qGqa}_3(s)&=&\frac{\qq\qGqb}{48\pi^2}\int_0^1d\alpha
\Bigg\{\frac{3m_Q^4(3-\alpha)}{(1-\alpha)\alpha^2}\delta'\left(s-\tilde{m}^2_Q\right)
\\&&+
\frac{m_Q^2(3\alpha^3-4\alpha^2-3\alpha+6)}{\alpha(1-\alpha)^2}\delta\left(s-\tilde{m}^2_Q\right)+(2\alpha+3)H\left(s-\tilde{m}^2_Q\right)
\Bigg\}\, . \label{SD13}
\end{eqnarray}
}
\item For the current $J_{4\mu}^{(\mathbf 1)}$
{\allowdisplaybreaks
\begin{eqnarray}
\nonumber
\rho^{{(\mathbf 1)}pert}_4(s)&=&\frac{1}{512\pi^6}\dab\Bigg\{\frac{9(1-\alpha-\beta)(1+\alpha+\beta)\f(s)^4}{4\alpha^3\beta^3}
\non&&-\frac{m_Q^2(1-\alpha-\beta)^2(5+\alpha+\beta)\f(s)^3}{\alpha^3\beta^3}\Bigg\}\, ,
\non
\rho^{{(\mathbf 1)}\qq}_4(s)&=&\frac{9m_Q\qq}{64\pi^4}\dab\frac{(1-\alpha-\beta)\f(s)\left[3m_Q^2(\alpha+\beta)-7\alpha\beta s\right]}{\alpha\beta^2}\, ,
\non
\rho^{{(\mathbf 1)}\GGa}_4(s)&=&\frac{\GGb}{1024\pi^6}\dab\Bigg\{\frac{3m_Q^2(1-\alpha-\beta)^2\left[m_Q^2(\alpha+\beta)-2\alpha\beta s\right]}{\alpha^3}
\non&&-\frac{m_Q^2(1-\alpha-\beta)^2(5+\alpha+\beta)\left\{3(\alpha^2+\beta^2)\f(s)+m_Q^2(\alpha^3+\beta^3)\right\}}{6\alpha^3\beta^3}
\non&&+\frac{(1-\alpha-\beta)\f(s)\left[m_Q^2(3+\alpha+\beta)+2\alpha\beta s\right]}{\alpha^2\beta}\Bigg\}\, ,
\non
\rho^{{(\mathbf 1)}\qGqa}_4(s)&=&-\frac{m_Q\qGqb}{64\pi^4}\dab
\non&&\Bigg\{\frac{(1-\alpha-\beta)\left[3m_Q^2(\alpha+\beta)-4\alpha\beta s\right]}{\beta^2}
+\frac{(2+7\alpha-2\beta)\left[3m_Q^2(\alpha+\beta)-5\alpha\beta s\right]}{2\alpha\beta}\Bigg\}\, ,
\non
\rho^{{(\mathbf 1)}\qq^2}_4(s)&=&\frac{5(2m_Q^2+s)\qq^2}{48\pi^2}\sqrt{1-4m_Q^2/s}\, ,
\non
\rho^{{(\mathbf 1)}\qq\qGqa}_4(s)&=&\frac{\qq\qGqb}{48\pi^2}\int_0^1d\alpha
\Bigg\{\frac{3m_Q^4(3-\alpha)}{(1-\alpha)\alpha^2}\delta'\left(s-\tilde{m}^2_Q\right)
\\&&+
\frac{m_Q^2(3\alpha^3-4\alpha^2-3\alpha+6)}{\alpha(1-\alpha)^2}\delta\left(s-\tilde{m}^2_Q\right)+(2\alpha+3)H\left(s-\tilde{m}^2_Q\right)
\Bigg\}\, . \label{SD14}
\end{eqnarray}
}
\item For the current $J_{2\mu}^{(\mathbf 8)}$
{\allowdisplaybreaks
\begin{eqnarray}
\nonumber
\rho^{{(\mathbf 8)}pert}_2(s)&=&\frac{1}{192\pi^6}\dab\frac{(1-\alpha-\beta)(1+\alpha+\beta)\f(s)^4}{\alpha^3\beta^3}\, ,
\non
\rho^{{(\mathbf 8)}\qq}_2(s)&=&\frac{m_Q\qq}{6\pi^4}\dab\frac{(1-\alpha-\beta)\f(s)\left[3m_Q^2(\alpha+\beta)-7\alpha\beta s\right]}{\alpha\beta^2}\, ,
\non
\rho^{{(\mathbf 8)}\GGa}_2(s)&=&-\frac{\GGb}{288\pi^6}\dab\Bigg\{\frac{7m_Q^2(1-\alpha-\beta)^2(5+\alpha+\beta)\f(s)}{96\alpha^2\beta^2}
\non&&+\frac{7m_Q^2(1+\alpha+\beta)\f(s)}{16\alpha\beta}-\frac{m_Q^2(1-\alpha-\beta)^2\left[m_Q^2(\alpha+\beta)-2\alpha\beta s\right]}{\alpha^3}
\non&&-\frac{(1-\alpha-\beta)\left[m_Q^2(3+\alpha+\beta)+2\alpha\beta s\right]\f(s)}{8\alpha^2\beta}\Bigg\}\, ,
\\
\rho^{{(\mathbf 8)}\qGqa}_2(s)&=&-\frac{m_Q\qGqb}{144\pi^4}\dab
\non&& \Bigg\{\frac{(1-\alpha-\beta)\left[6m_Q^2(\alpha+\beta)-11\alpha\beta s\right]}{\beta^2}
+\frac{17\left[3m_Q^2(\alpha+\beta)-5\alpha\beta s\right]}{2\beta}\Bigg\}\, ,
\non
\rho^{{(\mathbf 8)}\qq^2}_2(s)&=&\frac{4m_Q^2\qq^2}{9\pi^2}\sqrt{1-4m_Q^2/s}\, ,
\non
\rho^{{(\mathbf 8)}\qq\qGqa}_2(s)&=&\frac{\qq\qGqb}{108\pi^2}\int_0^1d\alpha
\Bigg\{\frac{24m_Q^4}{\alpha^2}\delta'\left(s-\tilde{m}^2_Q\right)+
\frac{m_Q^2(5\alpha^2-6\alpha+3)}{\alpha(1-\alpha)^2}\delta\left(s-\tilde{m}^2_Q\right)-\alpha H\left(s-\tilde{m}^2_Q\right)
\Bigg\}\, . \label{SD82}
\end{eqnarray}
}
\item For the current $J_{3\mu}^{(\mathbf 8)}$
{\allowdisplaybreaks
\begin{eqnarray}
\nonumber
\rho^{{(\mathbf 8)}pert}_3(s)&=&\frac{1}{16\pi^6}\dab\Bigg\{\frac{(1-\alpha-\beta)(1+\alpha+\beta)\f(s)^4}{4\alpha^3\beta^3}
\non&&-\frac{m_Q^2(1-\alpha-\beta)^2(5+\alpha+\beta)\f(s)^3}{9\alpha^3\beta^3}\Bigg\}\, ,
\non
\rho^{{(\mathbf 8)}\qq}_3(s)&=&-\frac{m_Q\qq}{2\pi^4}\dab\frac{(1-\alpha-\beta)\f(s)\left[3m_Q^2(\alpha+\beta)-7\alpha\beta s\right]}{\alpha\beta^2}\, ,
\non
\rho^{{(\mathbf 8)}\GGa}_3(s)&=&-\frac{\GGb}{96\pi^6}\dab\Bigg\{\frac{m_Q^2(1-\alpha-\beta)^2(5+\alpha+\beta)\left[m_Q^2(3\alpha+4\beta)-3\alpha\beta s\right]}{9\alpha^3\beta}
\non&&+\frac{7(1-\alpha-\beta)^2\f(s)\left[m_Q^2(13\alpha+13\beta+29)+12\alpha\beta s\right]}{288\alpha^2\beta^2}
\non&&-\frac{m_Q^2(1-\alpha-\beta)^2\left[m_Q^2(\alpha+\beta)-2\alpha\beta s\right]}{\alpha^3}
\non&&+\frac{7\f(s)\left[3m_Q^2(1+\alpha+\beta)+4\alpha\beta s\right]}{48\alpha\beta}
\non&&+\frac{(1-\alpha-\beta)\f(s)\left[m_Q^2(3+\alpha+\beta)+5\alpha\beta s\right]}{12\alpha^2\beta}\Bigg\}\, ,
\non
\rho^{{(\mathbf 8)}\qGqa}_3(s)&=&-\frac{m_Q\qGqb}{48\pi^4}\dab\Bigg\{
\frac{(1-\alpha-\beta)\left[3m_Q^2(\alpha+\beta)-5\alpha\beta s\right]}{\alpha\beta}
\non&&+\frac{(1-\alpha-\beta)\left[6m_Q^2(\alpha+\beta)-11\alpha\beta s\right]}{\beta^2}
-\frac{71m_Q^2(\alpha+\beta)-105\alpha\beta s}{2\beta}\Bigg\}\, ,
\non
\rho^{{(\mathbf 8)}\qq^2}_3(s)&=&\frac{10(2m_Q^2+s)\qq^2}{27\pi^2}\sqrt{1-4m_Q^2/s}\, ,
\non
\rho^{{(\mathbf 8)}\qq\qGqa}_3(s)&=&\frac{\qq\qGqb}{54\pi^2}\int_0^1d\alpha
\Bigg\{\frac{24m_Q^4(3-\alpha)}{2(1-\alpha)\alpha^2}\delta'\left(s-\tilde{m}^2_Q\right)
\\&&+
\frac{m_Q^2(24\alpha^3-91\alpha^2+54\alpha+9)}{2\alpha(1-\alpha)^2}\delta\left(s-\tilde{m}^2_Q\right)+(13\alpha+12)H\left(s-\tilde{m}^2_Q\right)
\Bigg\}\, . \label{SD83}
\end{eqnarray}
}
\item For the current $J_{4\mu}^{(\mathbf 8)}$
{\allowdisplaybreaks
\begin{eqnarray}
\nonumber
\rho^{{(\mathbf 8)}pert}_4(s)&=&\frac{1}{16\pi^6}\dab\Bigg\{\frac{(1-\alpha-\beta)(1+\alpha+\beta)\f(s)^4}{4\alpha^3\beta^3}
\non&&-\frac{m_Q^2(1-\alpha-\beta)^2(5+\alpha+\beta)\f(s)^3}{9\alpha^3\beta^3}\Bigg\}\, ,
\non
\rho^{{(\mathbf 8)}\qq}_4(s)&=&\frac{m_Q\qq}{2\pi^4}\dab\frac{(1-\alpha-\beta)\f(s)\left[3m_Q^2(\alpha+\beta)-7\alpha\beta s\right]}{\alpha\beta^2}\, ,
\non
\rho^{{(\mathbf 8)}\GGa}_4(s)&=&-\frac{\GGb}{96\pi^6}\dab\Bigg\{\frac{m_Q^2(1-\alpha-\beta)^2(5+\alpha+\beta)\left[m_Q^2(3\alpha+4\beta)-3\alpha\beta s\right]}{9\alpha^3\beta}
\non&&+\frac{7(1-\alpha-\beta)^2\f(s)\left[m_Q^2(13\alpha+13\beta+29)+12\alpha\beta s\right]}{288\alpha^2\beta^2}
\non&&-\frac{m_Q^2(1-\alpha-\beta)^2\left[m_Q^2(\alpha+\beta)-2\alpha\beta s\right]}{\alpha^3}
\non&&+\frac{7\f(s)\left[3m_Q^2(1+\alpha+\beta)+4\alpha\beta s\right]}{48\alpha\beta}
\non&&+\frac{(1-\alpha-\beta)\f(s)\left[m_Q^2(3+\alpha+\beta)+5\alpha\beta s\right]}{12\alpha^2\beta}\Bigg\}\, ,
\non
\rho^{{(\mathbf 8)}\qGqa}_4(s)&=&\frac{m_Q\qGqb}{48\pi^4}\dab\Bigg\{
\frac{(1-\alpha-\beta)\left[3m_Q^2(\alpha+\beta)-5\alpha\beta s\right]}{\alpha\beta}
\non&&+\frac{(1-\alpha-\beta)\left[6m_Q^2(\alpha+\beta)-11\alpha\beta s\right]}{\beta^2}
-\frac{71m_Q^2(\alpha+\beta)-105\alpha\beta s}{2\beta}\Bigg\}\, ,
\non
\rho^{{(\mathbf 8)}\qq^2}_4(s)&=&\frac{10(2m_Q^2+s)\qq^2}{27\pi^2}\sqrt{1-4m_Q^2/s}\, ,
\non
\rho^{{(\mathbf 8)}\qq\qGqa}_4(s)&=&\frac{\qq\qGqb}{54\pi^2}\int_0^1d\alpha
\Bigg\{\frac{24m_Q^4(3-\alpha)}{2(1-\alpha)\alpha^2}\delta'\left(s-\tilde{m}^2_Q\right)
\\&&+
\frac{m_Q^2(24\alpha^3-91\alpha^2+54\alpha+9)}{2\alpha(1-\alpha)^2}\delta\left(s-\tilde{m}^2_Q\right)+(13\alpha+12)H\left(s-\tilde{m}^2_Q\right)
\Bigg\}\, . \label{SD84}
\end{eqnarray}
}
\end{itemize}

\end{document}